\documentclass[aps,prl,twocolumn, showpacs]{revtex4-1}
\usepackage{amsmath}
\usepackage{amssymb}
\usepackage{graphics}

\newcommand\ro{\hat\rho}
\newcommand\Ho{\hat H}
\newcommand\Ao{\hat A}
\newcommand\bo{\hat b}
\newcommand\phio{\hat\phi}
\newcommand\Go{\hat G}
\newcommand\Lo{\hat L}
\newcommand\Lc{\mathcal{L}}
\newcommand\Mc{\mathcal{M}}
\newcommand\Xco{\hat{\mathcal{X}}}
\newcommand\om{\omega}
\newcommand\crm{\mathrm{c}}

\newcommand\dg{^\dagger}
\newcommand\pr{^\prime}
\newcommand\e{\mathrm{e}}
\newcommand\hc{\mathrm{h.c.}}
\newcommand\Tr{\mathrm{Tr}}

\newcommand\half{\frac{1}{2}}

\begin{document}
\title{General Non-Markovian structure of Gaussian Master and Stochastic Schr\"odinger Equations}   
\author{L. Di\'osi}
\email{diosi.lajos@wigner.mta.hu}
\affiliation{Wigner Research Center for Physics, H-1525 Budapest 114, P.O.Box 49, Hungary}
\author{L. Ferialdi}
\email{ferialdi@math.lmu.de}
\affiliation{Mathematisches Institut, Ludwig-Maximilians-Universit\"at, Theresienstr. 39, 80333 Munich.}
\date{\today}
\begin{abstract}
General open quantum systems display memory features, their master equations are non-Markovian. 
We show that the subclass of Gaussian non-Markovian open system dynamics 
is tractable in a depth similar to the Markovian class. The structure of master equations 
exhibits a transparent generalization of the Lindblad structure.  We find and parametrize the class of 
stochastic Schr\"odinger equations that unravel a given master equation,
such class was before known for Markovian systems only. 
We show that particular non-Markovian unravellings known in the literature are special cases of our class. 
\end{abstract}
\maketitle

The most successful and popular theories of open quantum systems request the Markovian approximation~\cite{BrePet02}. 
Markovian master equations (MMEs),  Markovian stochastic Schr\"odinger equations (MSSEs) \cite{WisDio01}, 
as well as Markovian quantum Langevin equations \cite{WisMil10} (not discussed here) proved to be powerful tools.
However, non-Markovian dynamics are acquiring a growing importance. 
Many ultra fast processes are non-Markovian, like e.g., light harvesting in photosynthesis~\cite{Thoetal}, 
ultra-fast chemical reactions~\cite{Pometal}, and photonic band gap materials~\cite{Garetal}.
The Markovian approximation is useful only when the system time scale is much slower than the one of the environment.  
When the time-scale of  the system is comparable to that of the bath (like in the examples listed above), 
the bath is not fast enough to go back to equilibrium, and some \lq\lq memory effects\rq\rq build up. 
The dynamics of the open system cannot be described by the  approximate memoryless MMEs and  MSSEs, 
but it requires a general non-Markovian description. 
The need of non-Markovian dynamics arose independently in foundations, too. 
Based on Strunz' work \cite{Str96},  the discovery of non-Markovian SSEs \cite{DioStr97} 
paved the way to relax the artificial markovianity of dynamical wave function collapse theories \cite{BasGir03}, 
leading to various non-Markovian models~\cite{BasetalFeretal}. 
One should bear in mind that 
non-Markovian is simply any dynamics which does not fall under the Markovian approximation, no further structure is implied.  
To further characterize a non-Markovian dynamics one must specialize the underlying structure.
The more particular this structure is, the less usable; the more general, the less analyzable in detail.

The aim of this Letter is to study a class of open systems which is sufficiently general, yet analytically tractable. 
We consider the most general non-Markovian structure for a Gaussian master equation (GME) 
and the related Gaussian stochastic Schr\"odinger equations (GSSEs). 
No restriction is made on the system's dynamics, just on the structure of the environment and the coupling. 
We shall see that our GME represents a very simple and intuitive generalization for non-Markovian dynamics of the Lindblad MME.  
The Gaussian option is a good compromise between generality and tractability.

{\it Markovian vs non-Markovian.}--
Generic MEs are given by the integral form
\begin{equation}\label{ME}
\ro_t=\Mc_t\ro_0
\end{equation}
where the evolution superoperator $\Mc_t$ is a trace preserving time-dependent completely positive (CP) map with $\Mc_0=1$.
In the special Markovian case the superoperator $\Mc_t$ can be written as
\begin{equation}\label{MMt}
\Mc_t=T\exp\left\{\int_0^t\!d\tau\Lc_t\right\},
\end{equation}
and the MME takes the differential form 
\begin{equation}\label{MME}
\frac{d\ro_t}{dt}=\Lc_t\ro_t\;,
\end{equation}
where the Lindblad superoperator $\Lc_t$ has a precise structure. 
If this form does not exist, we call the open system and its ME non-Markovian.
From now on we use the Heisenberg picture, 
where time-dependent operators $\Ao^j_t$ solve the Heisenberg equations with some system Hamiltonian $\Ho$. 
For later notational convenience, we introduce the left-right (LR) formalism \cite{Choetal85,Dio90,Dio93}, 
denoting by a subscript $L$ ($R$) the operators acting on $\ro$ from the left (right), 
e.g. $\Ao^k_{L}\Ao^j_{R}\ro=\Ao^k\ro\Ao^j$. With this notation the Lindblad superoperator reads
\begin{equation}\label{LindbladLR}
\Lc_t=D_{jk}(t)\left(\Ao^k_{tL}\Ao^j_{tR}-\half\Ao^j_{tL}\Ao^k_{tL}-\half\Ao^k_{tR}\Ao^j_{tR}\right)\,,
\end{equation}
where the Einstein summation over repeated latin indices has been assumed. 
$D_{jk}$ is an arbitrary  non-negative matrix, and $\Ao^j_t$ are Hermitian operators. 
In the special case of real coefficients $D_{jk}=D_{jk}^\star$ the MME describes fluctuations 
and decoherence without dissipation, so we call the MME non-dissipative; otherwise we call it dissipative. 

MMEs represent an approximation of the general non-Markovian ones: their simple mathematical structure allows for  
considerable insight and understanding \cite{BrePet02,WisDio01,WisMil10}. 
One important feature is that each MME can be identified as the reduced dynamics of the unitary dynamics 
of the system plus a suitably chosen `heat' bath.
As an alternative to MEs, SSEs are equally flexible tools.   
One can construct a SSE such that the mean of the random solutions $\vert\psi_t\rangle$ recovers the solution
\begin{equation}\label{Epsipsi}
\ro_t=\mathbb{E}\left[\;|\psi_t\rangle\langle\psi_t\vert\;\right]\;
\end{equation}
of the given ME. Then the SSE is said to unravel the ME. 
Each MME can be unravelled by an infinite number of equivalent MSSEs.
The full class of MSSEs unravelling the same MME is known and parametrized uniquely, 
identified as the modified unitary dynamics of the system under time-continuous quantum measurement \cite{WisDio01}.

Unlike MMEs, the features of generic MEs \eqref{ME} are little known since 
a general time-dependent CP-map $\Mc_t$ does not allow for much insight.
We have to study specific MEs: our choice will be the class of GMEs.

{\it Non-Markovian Gaussian ME.}--
We claim that the general form of a Gaussian evolution superoperator is
\begin{equation}\label{GMt}
\begin{array}{l}
\Mc_t=T\exp\left\{\int_0^t\!d\tau\!\!\int_0^t\!ds D_{jk}(\tau,s)\times\right.\\\;\;\;\;\;\;\;\;\;
\left.\times\left(\Ao^k_{sL}\Ao^j_{\tau R}\!-\!\theta_{\tau s}\Ao^j_{\tau L}\Ao^k_{s L}\!-\!\theta_{s\tau}\Ao^k_{sR}\Ao^j_{\tau R}\right)
\right\}\,,
\end{array}
\end{equation}
where the step function $\theta_{\tau s}$ is $1$ for $\tau>s$ while $0$ otherwise, 
and $T$ denotes time-ordering for both $L$ and $R$ operators~\cite{ordering}.
The kernel $D_{jk}(\tau,s)$ must be non-negative, we assume that it can be arbitrarily chosen otherwise.
Like in the Markovian case, for real kernel $D=D^\star$ we call the GME non-dissipative, and we call it dissipative otherwise.

It is easy to see that the Markovian evolution $\Mc_t$ of Eq.~\eqref{MMt} is a special case of Eq.~\eqref{GMt}.
Using the Lindblad form \eqref{LindbladLR}, the Markovian superoperator \eqref{MMt} reads:
\begin{equation}\label{GMtLindblad}
\begin{array}{l}
\Mc_t=T\exp\left\{\int_0^t\!d\tau D_{jk}(\tau)\times\right.\\\;\;\;\;\;\;\;\;\;
\left.\times\left(\Ao^k_{\tau L}\Ao^j_{\tau R}\!-\!\half\Ao^j_{\tau L}\Ao^k_{\tau L}\!-\!\half\Ao^k_{\tau R}\Ao^j_{\tau R}\right)
\right\}\;.
\end{array}
\end{equation}
This coincides exactly with the generic Gaussian $\Mc_t$ \eqref{GMt}, with
the special time-local choice of the kernel $D_{jk}(t,s)=D_{jk}(t)\delta(t-s)$.
Vice versa, one can formally obtain the GME from the MME by promoting the matrix $D_{jk}(t)$ to a double-time non-negative kernel, 
and adding a second integral over time. This represents an interesting insight into GMEs.

Since we claim that the GME is correct for all $D\geq0$, 
we have to prove that the superoperator $\Mc_t$ \eqref{GMt} is a trace-preserving CP-map for all $t>0$. 
The LR-formalism is surprisingly powerful to  prove that $\Mc_t$ preserves the trace. 
Let us introduce the notations $\Ao^j_\Delta=\Ao^j_L-\Ao^j_R$ and $\Ao^j_\crm=(\Ao^j_L+\Ao^j_R)/2$, cf. \cite{Choetal85,Dio93}. 
Then the exponent of \eqref{GMt} takes the following equivalent form:
\begin{equation}\label{GMtexp}
\begin{array}{l}
\Mc_t=T\exp\left\{-\half\int_0^t\!d\tau\!\!\int_0^t\!ds D_{jk}^\mathrm{(Re)}(\tau,s)\Ao^j_{\tau\Delta}\Ao^k_{s\Delta}-\right.\\\;\;\;\;\;\;\;\;\;
           \left.-2i\int_0^t\!d\tau\!\!\int_0^\tau\!ds D_{jk}^\mathrm{(Im)}(\tau,s)\Ao^j_{\tau\Delta}\Ao^k_{s\crm}
\right\}
\end{array}
\end{equation}
where $D^\mathrm{(Re)}=\mathrm{Re}D$ and $D^\mathrm{(Im)}=\mathrm{Im}D$ are real symmetric and antisymmetric kernels respectively.
Obviously $\Ao^j_{\tau\Delta}$, $\Ao^k_{s\Delta}$ represent commutators, which do never make the trace of $\Mc_t\ro_0$ change. 
The superoperators $\Ao^k_{s\crm}$ represent anti-commutators, which might make  $\Mc_t$ change the trace, 
but they do not since they always appear in combination like $\Ao^j_{\tau\Delta}\Ao^k_{s\crm}$: 
the anti-commutation is always followed by at least one commutation. 
Hence the map $\Mc_t$ preserves the trace~of~$\ro_t$. That $\Mc_t$ is CP, will be proved later.

\label{MEreddyn}
We might prove that $\Mc_t$ is both trace-preserving and CP, 
by identifying $\Mc_t$ as the reduced dynamics of the unitary dynamics of the system plus a suitably chosen bosonic `heat' bath.
Assume the following Hamiltonian bilinear coupling between our system and the bath:
\begin{equation}\label{coupling}
\Ao^j_t\phio_j(t)\,,
\end{equation} 
where $\phio_j(t)$ are bosonic fields of the bath in interaction picture. 
We assume that there exists a central Gaussian bath initial state $\ro_B$ such that
the fields correlation function coincides with the kernel $D$ in 
Eq.~\eqref{GMt} of our GME:
\begin{equation}\label{phiphiD}
\Tr_B\left[\phio_j(\tau)\phio_k(s)\ro_B\right]=D_{jk}(\tau,s)\;.
\end{equation}
This assumption will be discussed later.
The system-bath state $\ro_{SB}$ evolves with the von Neumann equation
\begin{equation}\label{roSBdot}
\frac{d\ro_{SBt}}{dt}=-i\left(\!\Ao^j_{t L}\phio_{jL}(t)\!-\!\Ao^j_{t R}\phio_{jR}(t)\!\right)\!\ro_{SBt}.
\end{equation}
For an uncorrelated initial state $\ro_0\ro_B$, 
one can write the reduced dynamics into the form \eqref{ME}:
\begin{equation}
\begin{array}{l}\label{RedDyn}
\Mc_t\ro_0=\\
\Tr_B \left[T\exp\left\{-i\int_0^t\!d\tau
                      \left(\Ao^j_{\tau L}\phio_{jL}(\tau)-\Ao^j_{\tau R}\phio_{jR}(\tau)\right)\right\}\ro_0\ro_B\right]\\
\equiv \Tr_B\left[T\exp(-i\Xco)\ro_0\ro_B\right].
\end{array}
\end{equation}
We apply the following identity valid in Gaussian state $\ro_B$ (see e.g. \cite{Choetal85,Dio90,Dio93} 
and references therein):
\begin{equation}\label{Wick}
\Tr_B\left[T\exp(-i\Xco)\ro_B\right]=T\exp\left(-\half\Tr_B[T\Xco^2\ro_B]\right)
\end{equation}
where $\Xco$ is an arbitrary linear functional of the bosonic fields.
Identifying  $\Xco$ as the integral  in \eqref{RedDyn}, one finds:
\begin{equation}
\begin{array}{l}
\Tr_B[T\Xco^2\ro_B]=-2\int_0^t\!d\tau\!\int_0^t\!ds D_{jk}(\tau,s)\times\\
\;\;\;\;\;\;\;\;\;
\times\left(\Ao^k_{sL}\Ao^j_{\tau R}\!-\!\theta_{\tau s}\Ao^j_{\tau L}\Ao^k_{sL}\!-\!\theta_{s\tau}\Ao^k_{sR}\Ao^j_{\tau R}\right)
\end{array}
\end{equation}
upto operator ordering. 
Using this result  in Eqs.~(\ref{RedDyn}-\ref{Wick}), one obtains Eq.~\eqref{GMt} for $\Mc_t$, as expected.
Let us come back to the assumption \eqref{phiphiD}.  
We confirm it for the special case of time-translation invariant kernels:
\begin{equation}\label{DFourier}
D_{jk}(\tau,s)=\int \e^{-i\om(\tau-s)}{\Tilde D}_{jk}(\om) \frac{d\om}{2\pi},
\end{equation}
${\Tilde D}_{jk}(\om)$ is an arbitrary non-negative Hermitian matrix, function of $\om\in(-\infty,\infty)$.
Let the `heat' bath consist of a continuum of harmonic oscillators of both positive and negative 
frequencies $\om$ \cite{GarCol85}.  Let the fields interacting with our system be Hermitian linear combinations of 
the free bosonic modes of the bath:
\begin{equation}\label{field}
\phio_j(t)=\int\kappa_j^l(\om)\bo_{l\om}\e^{-i\om t}d\om+\hc
\end{equation}
with $[\bo_{j\om},\bo\dg_{k\om\pr}]=\delta_{jk}\delta(\om-\om\pr)$. 
Assume the vacuum state for $\ro_B$, defined by $\bo_{k\om}\ro_B\equiv0$. 
The fields correlation is:
\begin{equation}\label{phiphi}
\Tr\left[\phio_j(\tau)\phio_k(s)\ro_B\right]
 =\int\kappa_j^l(\om)\kappa_k^{l\ast}(\om)\e^{-i\om(\tau-s)}d\om.
\end{equation}
Comparing this result with Eq.~\eqref{DFourier}, we see that the desired relationship \eqref{phiphiD}
is satisfied if ${\Tilde D}_{jk}(\om)=2\pi\kappa_j^l(\om)\kappa_k^{l\ast}(\om)$ which 
can always be ensured by the choice of the complex coefficients $\kappa_j^l(\om)$.

We have shown the bath correlation functions exhaust all time-translation invariant non-negative kernels.  
Hence the Gaussian $\Mc_t$ is trace-preserving CP-map contributing to a correct GME for all non-negative kernels $D$
that are time-translation invariant.
The correctness of our GME for all non-negative kernels $D$ will be proved  in an alternative way below. 

{\it Non-Markovian Gaussian SSE.}--
We begin with the simple non-dissipative case $D=D^\star$,
we show that the GME \eqref{GMt} is equivalent, in the sense of unravelling \eqref{Epsipsi}, 
with the average unitary dynamics of the system in colored classical Gaussian real noises $\phi_j(t)$. 
Consider the bilinear coupling $\Ao^j_t\phi_j(t)$,
and choose the real (non-dissipative) correlation of the noises such that $\mathbb{E}[\phi_j(\tau)\phi_k(s)]=D_{jk}(\tau,s)$.
Then the wave function evolves according to the following GSSE: 
\begin{equation}\label{SSEuni}
\frac{d|\psi_t\rangle}{dt}=-i\Ao^j_t\phi_j(t)|\psi_t\rangle\,.
\end{equation}
The solution can be written in the compact form
$|\psi_t\rangle=G_t[\phi]|\psi_0\rangle$ by introducing the Green-operator
\begin{equation}\label{Gt}
\Go_t[\phi]=T \exp\left[-i\int_0^tds \Ao_s^j \phi_j(s)\right]\;.
\end{equation}
Thereby, one finds that $\ro_t$ evolves according to Eq.~\eqref{ME} with superoperator
\begin{equation}\label{Mtphi}
\Mc_t=\mathbb{E}\left\{T \exp\left[-i\int_0^tds (\Ao_{sL}^j-\Ao_{sR}^j) \phi_j(s)\right]\right\}\,.
\end{equation}
Performing the stochastic average, since the rule  \eqref{Wick} applies invariably
if $\Tr_B[\dots\ro_B]$ is replaced by $\mathbb{E}[\dots]$, one recovers Eq.~\eqref{GMt}.

This proves that a non-dissipative GME is equivalent to the averaged unitary dynamics with real colored noise.
The corresponding GSSE \eqref{SSEuni} represents one of the (infinite many) possible unravellings of the non-dissipative GME. 
Of course it also means the superoperator $\Mc_t$ \eqref{GMt} is trace-preserving CP-map for all real kernels $D=D^\star$.

Unlike the non-dissipative GME, a dissipative GME (i.e., with complex kernel $D\neq D^\star$) cannot be  unravelled by the 
simple GSSE \eqref{SSEuni}. 
One has to relax the unitarity
of the dynamics, letting the Hamiltonian coupling be non-Hermitian.
Still, we start from the old coupling $\Ao^j_t\phi_j(t)$, GSSE \eqref{SSEuni}, and Green-operator \eqref{Gt}.    
Now $\phi_j(t)$ are complex colored noises,
to allow for a complex (dissipative) correlation, that we set equal to the kernel $D$ of the GME:
\begin{equation}\label{corrherm}
\mathbb{E}\left[\phi_j^\ast(\tau)\phi_k(s)\right]=D_{jk}(\tau,s)\;.
\end{equation}
There is a further independent complex symmetric (non-Hermitian) correlation:   
\begin{equation}\label{corrsymm}
\mathbb{E}\left[\phi_j(\tau)\phi_k(s)\right]=S_{jk}(\tau,s)\,,
\end{equation}
which is only constrained by positivity of the full correlation kernel
\begin{equation}\label{corrfull}
\left(\begin{array}{ll}D&S\\S^\ast&D^\ast\end{array}\right)\geq0.
\end{equation}
The Green-operator of Eq.~\eqref{Gt} is not unitary in the dissipative case. 
It does not preserve the normalization of $|\psi_t\rangle$, 
but the crucial unravelling condition \eqref{Epsipsi} must remain valid. 
Hence we have to check whether
\begin{equation}\label{EGroG}
\ro_t=\Mc_t\ro_0=\mathbb{E}\left\{\Go_t[\phi]\ro_0\Go_t\dg[\phi^\ast]\right\}
\end{equation}
yields the solution $\ro_t$ with the Gaussian evolution superoperator \eqref{GMt}.
Insert Eq.~\eqref{Gt} and evaluate the stochastic mean. Due to the symmetry of the
kernel $S$, the resulting superoperator can be written as follows: 
\begin{equation}\label{noGMt}
\begin{array}{l}
\Mc_t=T\exp\left\{\int_0^t\!d\tau\int_0^t\!ds\left(D_{\!jk}(\!\tau\!,\!s)\Ao^k_{s\!L}\Ao^j_{\tau\!R}\!\!\right.\right.\\
\left.\left.-
                 \!\theta_{\tau s}S_{\!jk}     (\!\tau\!,\!s)\Ao^j_{\tau\!L}\Ao^k_{s\!L}\!\!-                                   
                  \!\theta_{s \tau}S_{\!jk}^\ast(\!\tau\!,\!s)\Ao^k_{s\!R}\Ao^j_{\tau\!R}\right)\right\}\;.
\end{array}
\end{equation}
This is clearly different from the desired form \eqref{GMt}. The LR term is correct but the kernels of
the LL and RR terms are $S_{jk}(\tau,s)$ and $S_{jk}^\ast(\tau,s)$ respectively, instead of the correct 
$D_{jk}(\tau,s)$. 
However, one can correct the Green-operator \eqref{Gt}
by adding suitable counter-terms:
\begin{equation}\label{Gtcorrect}
\begin{array}{l}
\Go_t[\phi]\!=\!T\!\left\{\!\exp\!\left(\!-i\!\int_0^t\!\!ds \Ao_s^j\phi_j(s)\!\right.\right.\;\;\;\;\;\\
                \left.\left.\!-\!\int_0^t\!d\tau\!\!\!\int_0^t\!ds \theta_{\tau s}
                                 [D_{\!jk}\!(\!\tau\!,\!s)\!-\!S_{\!jk}\!(\!\tau\!,\!s)]\!\Ao^j_{\tau}\Ao^k_{s}\!\right)\!\right\}\;.
\end{array}
\end{equation}
If we re-evaluate Eq.~\eqref{EGroG} with the new Green-operator above, we get exactly the 
desired superoperator \eqref{GMt}.

Following the method of \cite{DioStr97}, we read out the GSSE from the solutions $\Go_t[\phi]\vert\psi_0\rangle$:
\begin{equation}\label{SSEdiss}
\frac{d|\psi_t\rangle}{dt}=-i\Ao^j_t
                           \!\left(\!\phi_j(t)\!+\!\int_0^t\!d\tau 
                                 [D_{\!jk}\!(\!t\!,\!s)\!-\!S_{\!jk}\!(\!t\!,\!s)]\frac{\delta}{\delta\phi_k(\tau)}\!\right)
\!|\psi_t\rangle\;
\end{equation}
This form is valid if the kernels $D$ and $S$ have no equal-time finite-measure singularity. 
On the contrary, in the Markovian case,  when  the kernel $D$ of the GME is time-local 
(yielding to the MME), the symmetric kernel must also be reduced to a time-local one: $S_{jk}(t,s)=S_{jk}(t)\delta(t-s)$.
Using the Markovian kernels in the Green-operator, we read-out the following MSSE: 
\begin{equation}\label{MSSEdiss}
\frac{d|\psi_t\rangle}{dt}=\left(-i\Ao^j_t\phi_j(t)-\half [D_{jk}(t)-S_{jk}(t)]\Ao^j_t\Ao^k_t\right)|\psi_t\rangle\;,
\end{equation}
the symmetric matrix $S_{jk}(t)$ yields the parameters of the different MSSEs unravelling the same MME, 
in accordance with the Markovian theory \cite{WisDio01}.

The result \eqref{SSEdiss} is the most general non-Markovian GSSE in interaction picture, to unravel a general GME. 
Similarly to the MMEs, there is an infinite variety of GSSEs for each GME. The symmetric kernel $S$ represents 
the continuum many free parameters. Note, finally, 
that the very existence of our unravellings does prove that $\Mc_t$ is CP-map because it is of the Kraus form \eqref{EGroG}.  
Since we previously proved that $\Mc_t$ is trace-preserving, 
the proof of correctness of our GME with any non-negative kernel $D$ is complete. 

We now show that by exploiting the freedom of tuning $S$ we choose specific noises in Eq.~\eqref{SSEdiss} 
and  we can recover all previously  known SSEs.
If $\phi$ is complex Hermitian noise, then $S=0$ and Eq.~\eqref{SSEdiss} reduces to the quantum state diffusion SSE 
first proposed in~\cite{DioStr97}:
\begin{equation}\label{SSEqsd}
\frac{d|\psi_t\rangle}{dt}=-i\Ao_t^j
                           \left(\phi_j(t)+\int_0^t\!ds 
                                 D_{jk}(t,s)\frac{\delta}{\delta\phi_k(s)}\right)
|\psi_t\rangle\;.
\end{equation}
In standard non-Markovian quantum state diffusion SSE the operators $\Ao^j$ are not necessarily Hermitian. 
We show on a simple example how our GMEs yield this general case as well. 
Suppose we have just two Hermitian operators $\Ao^1,\Ao^2$, 
assume a degenerate kernel satisfying $D_{11}=D_{22}=D$ and $D_{12}=D_{21}^{\star}=-iD$. This means we have two perfect correlated
Hermitian noises satisfying $\phi_1=i\phi_2=\phi$. Applying this setting in the GSSE \eqref{SSEdiss}, for $\Lo=\Ao^1+i\Ao^2$ we get:
\begin{equation}\label{SSEqsdNonherm}
\frac{d|\psi_t\rangle}{dt}=\left(-i\Lo_t^\dagger\phi(t)-i\Lo_t\int_0^t ds D(t,s)\frac{\delta}{\delta\phi(s)}\right)\;
|\psi_t\rangle\;,
\end{equation}
which really generalizes \eqref{SSEqsd} for non-Hermitian operators. 
These equations have constantly been studied and applied in different contexts,  
from quantum foundations to quantum chemistry~\cite{DioGisStr98etc}.  

The following cases represent two extreme GSSEs unraveling the same non-dissipative GME. 
The first special case corresponds to unitary evolution while the second to a process of dynamical collapse. 
This duality was elucidated for an MME with two different MSSEs longtime ago \cite{Dio88},
here we are going to point out the same duality for our non-Markovian open systems.
If we choose $S=D$, hence $\phi$ is a real noise, 
and we recover the non-dissipative unitary GSSE \eqref{SSEuni} previously discussed.
On the contrary, if we set $S=-D$, hence $\phi$ is purely imaginary, and we obtain the collapse SSE \cite{BasGhi02GamWis02}:
\begin{equation}\label{SSEcoll}
\frac{d|\psi_t\rangle}{dt}=-i\Ao_t^j
                           \left(\phi_j(t)+2\int_0^t\!ds 
                                 D_{jk}(t,s)\frac{\delta}{\delta\phi_k(s)}\right)
|\psi_t\rangle\;.
\end{equation}
These equations describe the evolution of a wave function subject to random unsharp collapses on the eigenstates of $\Ao^j$. 
Of particular interest is the case when $\Ao$ is the position operator, 
which has been subject of thorough study  ~\cite{BasetalFeretal,Dio08WisGam08} in context of quantum foundations.

{\it Markovian limit.}--
We have already shown that, by choosing local kernels $D$ and $S$, one recovers the known MMEs, 
both in the dissipative and non-dissipative cases.
More than that, a general GME may possess a Markovian limit in some particular regimes, 
recovering the well known Lindblad equation~\eqref{LindbladLR}. 
A precise way to perform the Markovian limit of open quantum systems in stationary baths is the so-called 
rotating wave approximation \cite{BrePet02}, provided that the Fourier transform $\Ao^j_\om$ of $\Ao^j_t$ is discrete, for all $j$,
i.e.: $\Ao^j_t=\sum_\om\Ao^j_\om\e^{-i\om t}$. Applying rotating wave approximation to our superoperator $\Mc_t$~\eqref{GMt}, 
one obtains a stationary Lindblad superoperator.
We might apply this approximation perturbatively to the system+bath unitary dynamics~\eqref{roSBdot}, as is usually done. 
Here we apply it directly to Gaussian open system dynamics, i.e., 
to the exponent in the Gaussian superoperator $\Mc_t$ \eqref{GMt}. 
Comparing the result with \eqref{MMt},  we get the following stationary Lindblad superoperator:
\begin{equation}\label{MMErotwave}
\Lc\!=\!\sum_\om\!{\Tilde D}_{jk}\!(\om)\!\left(\!\Ao^k_{\om L}\Ao^{j\dagger}_{\om R}
                                          \!-\!\half\Ao^{j\dagger}_{\om L}\Ao^k_{\om L}\!-\!\half\Ao^k_{\om R}\Ao^{j\dagger}_{\om R}\!\right)\;.
\end{equation}
If one considers the dissipative dynamics of a free particle, the rotating wave approximation cannot be used. 
In this case, for a high-temperature heat bath, one can approximate the kernel by a quasi-time-local expression.
Following some heuristic steps, the calculation leads to a Lindblad MME of quantum Brownian motion \cite{Dio93}.

{\it Summary.}-- 
We analyzed the class of GMEs which is suitably general yet analytically tractable. 
We generalized the fundamental features of the well-known and well-tractable Lindblad MMEs for the proposed non-Markovian GMEs. 
Interestingly, the  evolution superoperator \eqref{GMt} of the GME can be formally generalized 
from the Lindblad structure \eqref{LindbladLR}, 
by promoting the Lindblad matrix $D_{jk}(t)$ to a double-time kernel $D_{jk}(\tau,s)$. 
This relationship gives a concrete insight into the way the GMEs work compared to the much studied and simpler MMEs. 
The GME is completely determined by a set of  Heisenberg operators  $\Ao^j_t$ and by the non-negative kernel $D$. 
It was known before, e.g. from \cite{CalLeg83,Dio90}, that the structures like our GME are reduced dynamics in bosonic reservoirs. 
Remarkably enough, we found it non-trivial whether all GMEs are reduced dynamics, 
the proof exists for  time-translation invariant kernels only. 
One major result is that we proved the correctness of the GMEs for all non-negative kernels $D$ 
whether or not the embedding heat bath exists.
Furthermore, we have generalized the classification of all stochastic unravellings for the non-Markovian GMEs. 
For a given GME, all GSSEs are uniquely parametrized by a certain symmetric kernel $S_{jk}(\tau,s)$, 
in full analogy with the corresponding symmetric matrix that parametrizes  the Markovian SSEs in \cite{WisDio01}. 
We showed that all non-Markovian SSEs known before  are specific cases of our  GSSEs, 
corresponding to various choices of the symmetric kernel $S$.

\phantom{}
The work of LF was supported by the Marie Curie Fellowship PIEF-GA-2012-328600. LF thanks A. Smirne for many useful conversations.
LD was supported by EU COST Action M1006.


\begin{thebibliography}{99}
\bibitem{ordering} One might question the unicity of operator ordering in the above expression \eqref{GMt}, 
since, e.g.,  $T\Ao^j_\tau\Ao^k_\tau$ remains undefined for $[\Ao^j_\tau,\Ao^k_\tau]\neq0$. 
However, the ambiguity is irrelevant as long as the kernel $D$ does not contain 
a singularity of finite integral measure at $\tau=s$. 
If it does, still we get a unique Markovian contribution, as shown in the main text.
\end{thebibliography}
\end{document}